\documentclass[times, twoside]{rxiv_maker_style}
\usepackage{blindtext} 
\usepackage{float} 

\renewcommand{\today}{2025-06-20}

\leadauthor{Saraiva}

\begin{document}
\title{{\color{red}R}$\chi$iv-Maker: an automated template engine for streamlined scientific publications}

\shorttitle{{\color{red}R}$\chi$iv-Maker}

\author[1,\Letter]{Bruno M. Saraiva}
\author[1]{Rita Carlota}
\author[1]{António D. Brito}
\author[2,3,4]{Iván Hidalgo-Cenalmor}
\author[2,3,4,\Letter]{Guillaume Jacquemet}
\author[1,\Letter]{Ricardo Henriques}
\affil[1]{Instituto de Tecnologia Química e Biológica António Xavier, Universidade Nova de Lisboa, Oeiras, Portugal}
\affil[2]{Turku Bioscience Centre, University of Turku and Åbo Akademi University, Turku, Finland}
\affil[3]{Faculty of Science and Engineering, Cell Biology, Åbo Akademi University, Turku, Finland}
\affil[4]{InFLAMES Research Flagship Center, University of Turku, Turku, Finland}

\maketitle

\begin{abstract}

The rapid growth of preprint servers has accelerated scientific dissemination but has also shifted the technical burden of manuscript preparation to authors. This challenge is particularly acute in computational research, where manuscripts must remain synchronised with evolving data and code. We present Rxiv-Maker, a framework that resolves this by converting simple Markdown files into professionally typeset, publication-ready PDFs. Its core feature is the ability to execute embedded code, creating a self-updating manuscript where figures and statistical values are generated directly from source data during compilation. This ensures that the final document is always current and fully reproducible. By building on standard tools like Git and Visual Studio (VS) Code, Rxiv-Maker offers a transparent, collaborative authoring workflow that applies software-engineering practices to academic writing.

\end{abstract}

\begin{keywords}
article template | scientific publishing | preprints
\end{keywords}

\begin{corrauthor}
(B. M. Saraiva) bsaraiva\at itqb.unl.pt; 
(G. Jacquemet) guillaume.jacquemet\at abo.fi; 
(R. Henriques) r.henriques\at itqb.unl.pt
\end{corrauthor}

\section*{Introduction}

The rise of preprint servers has reshaped scientific publishing, allowing rapid dissemination of research findings across numerous platforms \cite{beck2020,levchenk2024,Fraser2020_preprint_growth} (Fig. \ref{sfig:arxiv_growth}, Fig. \ref{sfig:preprint_trends}). This acceleration, however, has transferred the complex task of typesetting from publishers to researchers \cite{Vale2015_preprints,Tenant2016_academic_publishing,lin2020}. For those of us in computational fields, this is compounded by a more pressing challenge: ensuring the manuscript remains perfectly synchronised with our data and analysis code. We have all faced the tedious and error-prone task of manually updating a p-value or a sample size in the text after re-running an analysis \cite{perkel2022}. This disconnect between the research and the report is a critical issue, particularly in disciplines like bioimage analysis, where our findings are built upon complex computational pipelines \cite{biaflows2024,dl4miceverywhere2024}.

\begin{figure}[t]
\centering
\makebox[\linewidth][c]{
  \includegraphics[width=\linewidth,keepaspectratio,draft=false]{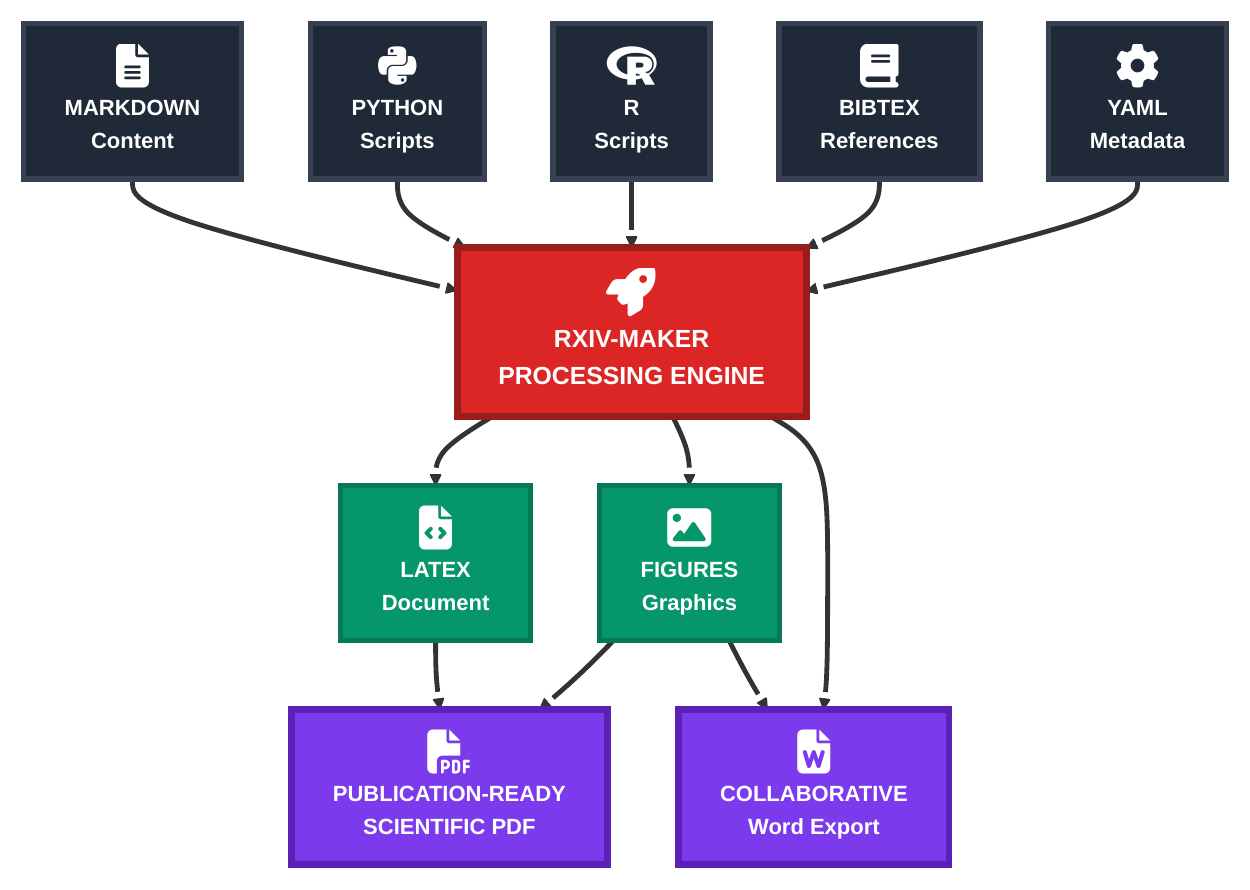}
}
\begingroup\captionsetup{width=\linewidth,singlelinecheck=false,justification=justified}\setlength{\abovecaptionskip}{6pt}\setlength{\belowcaptionskip}{6pt}\ifdefined\justifying\justifying\fi\caption{\textbf{System Architecture.} Rxiv-Maker integrates Markdown content, YAML metadata, executable scripts, and bibliographies through a processing engine that combines local execution with LaTeX PDF compilation or Word (.docx) generation to produce publication-ready documents.}\label{fig:system_diagram}\endgroup
\end{figure}

\begin{figure*}[t]
\centering
\makebox[\textwidth][c]{
  \includegraphics[width=\textwidth,keepaspectratio,draft=false]{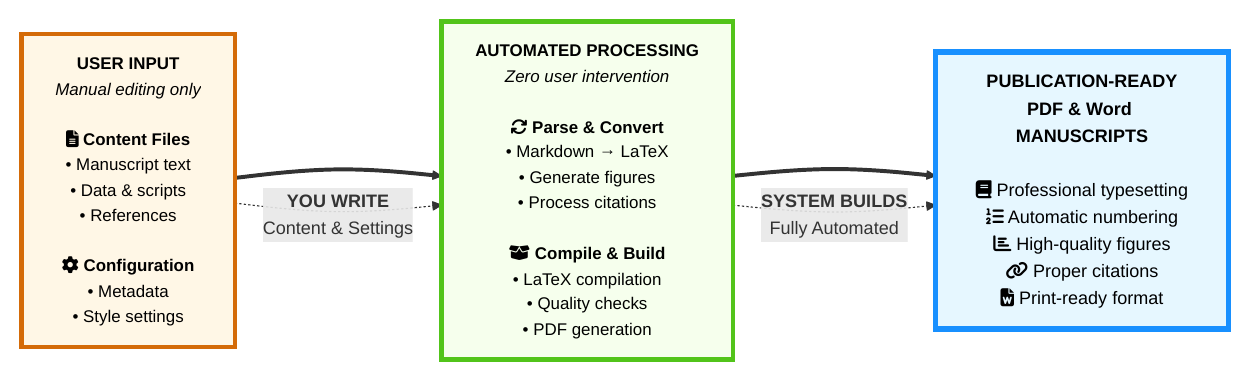}
}
\begingroup\captionsetup{width=\textwidth,singlelinecheck=false,justification=justified}\setlength{\abovecaptionskip}{6pt}\setlength{\belowcaptionskip}{6pt}\ifdefined\justifying\justifying\fi\caption{\textbf{Processing Pipeline.} User-provided content (left) undergoes automated processing (right), including parsing, script execution, document compilation (PDF/Word), and final output generation.}\label{fig:workflow}\endgroup
\end{figure*}

To address these challenges, we developed Rxiv-Maker, a framework that simplifies the creation of scientific manuscripts. It produces publication-ready documents from simple Markdown, using LaTeX for typesetting or exporting to Microsoft Word, without asking the author to manage formatting by hand. By executing embedded Python or R scripts during compilation, Rxiv-Maker turns the document into a self-updating manuscript that links the data directly to the final publication, removes manual transcription errors, and keeps the text synchronised with the latest results. It runs locally for a fast, responsive authoring experience and works with version control systems like Git for transparency and collaboration \cite{Ram2013_git_science,Perez-Riverol2016_github_bioinformatics}. The overall architecture and processing pipeline are illustrated in Fig. \ref{fig:system_diagram} and Fig. \ref{fig:workflow}, respectively. A tutorial is provided to guide new users through their first manuscript (\ref{snote:getting_started}).

\section*{Results}

\subsubsection{From Static Text to Self-Updating Manuscripts}

\begin{figure}[b]
\centering
\makebox[\linewidth][c]{
  \includegraphics[width=\linewidth,keepaspectratio,draft=false]{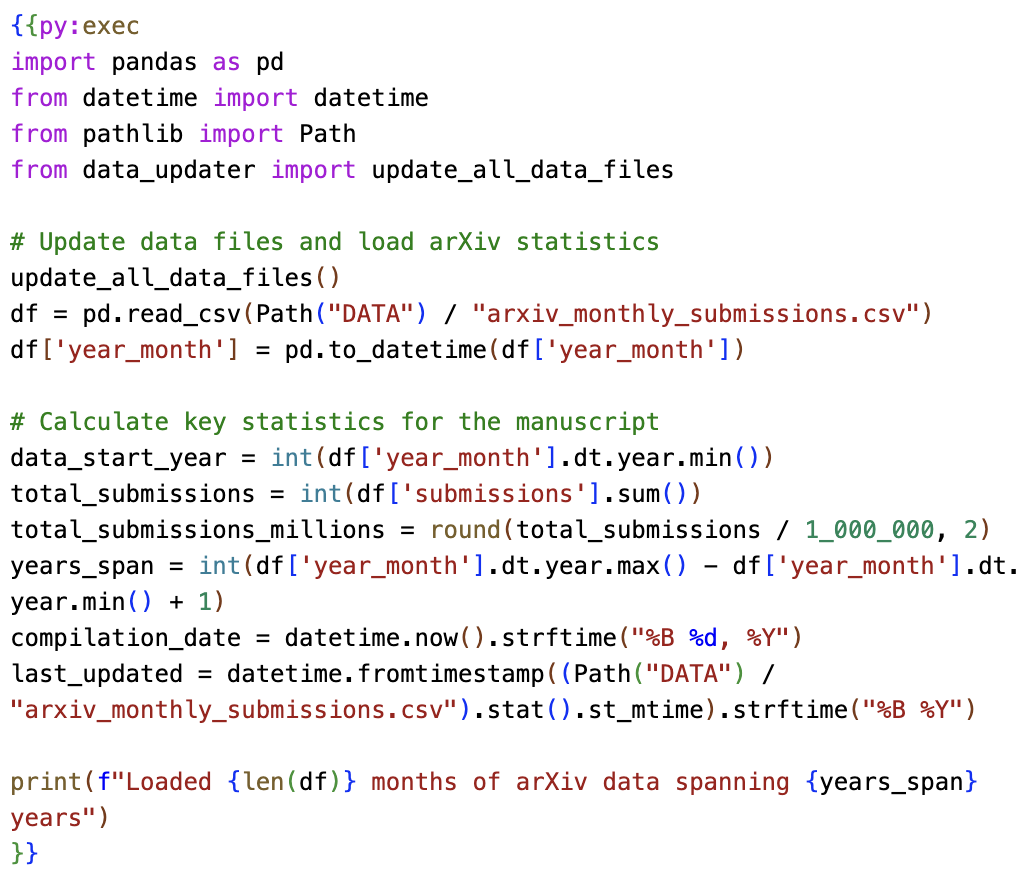}
}
\begingroup\captionsetup{width=\linewidth,singlelinecheck=false,justification=justified}\setlength{\abovecaptionskip}{6pt}\setlength{\belowcaptionskip}{6pt}\ifdefined\justifying\justifying\fi\caption{\textbf{Embedded Python for dynamic content.} This figure shows a Python script embedded in the manuscript's source code. The script is executed at build time to compute data attributes (e.g., total submissions and span of years) and inject the resulting values directly into the text. This process, rendered here with syntax highlighting from the Rxiv-Maker VS Code extension, eliminates manual transcription errors and ensures the text remains synchronised with the source data.}\label{fig:python_snippet}\endgroup
\end{figure}

A core feature of Rxiv-Maker is its ability to create self-updating documents. By embedding executable code snippets directly within the Markdown source, the manuscript changes from a static report to a document that regenerates its own results. This functionality enables the direct execution of analysis scripts during compilation, with the results (whether statistical values, tables, or figures) directly injected into the text. This process eliminates the possibility of transcription errors when, for example, copying a value from a terminal or another program into the manuscript.
We demonstrate this with a practical example from this very paper. The Python code shown in Fig. \ref{fig:python_snippet} is embedded in the source of our manuscript. At build time, it retrieves the latest public data on arXiv submissions, calculates key statistics, and inserts them into the text. This ensures that our discussion of preprint growth is always based on the most current evidence available. The analysis reveals that, since 1991, a total of 3.1 million submissions have been made over 36 years (Fig. \ref{sfig:arxiv_growth}). These are not static numbers; they were computed when this document was compiled on July 09, 2026, using data updated in July 2026. This capability extends to all forms of programmatic figure generation, as detailed in \ref{snote:programmatic_figures}, ensuring that every visual and numerical claim is reproducibly generated.

\subsubsection{Multi-Format Production: From PDF to Word}

Rxiv-Maker allows researchers to achieve the typographic quality of LaTeX while writing in simple, intuitive Markdown. This works through a multi-pass translator that converts Markdown into a restricted subset of LaTeX. A single-pass conversion would be too fragile for academic documents, where syntax for mathematics, citations, and cross-references must be preserved. The translator first identifies and protects delicate elements: mathematical expressions (\ref{snote:mathematical_formulas}), code blocks, and citation keys. Later passes normalise document structure and convert Rxiv-Maker's extended Markdown syntax into the corresponding LaTeX commands.

Rxiv-Maker also exports beyond PDF, since co-authors or journals may need editable formats. A dedicated Word export engine (built on \texttt{python-docx}) maps the same Markdown source to \texttt{.docx}, preserving figure numbering, tables, citations, and equations for exchange with collaborators. To support review, the \texttt{track-changes} feature uses Git history to generate annotated PDFs (via \texttt{latexdiff}) that show text changes between manuscript versions or against specific milestones (e.g., submission vs. revision).

The system handles complex nested formatting, inline DOI resolution, and transforms image syntax into floating figure environments, automatically handling captions, labels, and layout parameters. This enables authors to specify figure widths or positions using simple options in Markdown, without needing to modify LaTeX code. The system supports a range of figure generation methods, from static images to script-based visualisations and text-based diagrams written with Mermaid (Table \ref{stable:figure_formats}), so all visual elements are typeset consistently. For more advanced formatting needs, including complex tables, users can directly inject LaTeX code using \texttt{\detokenize{{{tex:...}}}} blocks. A complete syntax reference and example are detailed in \ref{snote:markdown-syntax}.

\subsubsection{An Integrated and Rapid Authoring Experience}

We designed Rxiv-Maker for a fast, responsive writing experience. It runs as a local-first command-line tool, so compilation is quick and authors get immediate feedback on their changes. A caching system tracks which figures and results are already built by computing a checksum from each figure's source code and data dependencies; a script is only re-executed when that checksum changes, saving time during compilation without compromising reproducibility (\ref{snote:caching_validation}).
To lower the barrier to entry, we developed a companion extension for Visual Studio Code (Fig. \ref{sfig:vscode_extension}). It provides syntax highlighting for the extended Markdown, autocompletion for citation keys and cross-references, and real-time validation of the manuscript configuration, keeping authoring and validation in one familiar environment so researchers can focus on content instead of typesetting.

\subsubsection{Designed for Transparent and Collaborative Science}

Reproducibility and collaboration are at the heart of Rxiv-Maker. By design, the entire manuscript is structured to be managed with Git, the version control system that is standard in software development. This includes the text, code, data, and configuration. This treats the manuscript as a complete, self-contained project where every change is tracked, attributed, and auditable. This aligns perfectly with the principles of open science, creating a transparent history of the research from its inception to the final publication.
The framework's command-line interface includes a powerful \texttt{rxiv validate} command that serves as a quality control gatekeeper. It performs a series of checks, such as verifying that all figures are present, all cross-references are valid, and all bibliographic entries are correctly formatted. This validation can be integrated into automated workflows, for instance, as a pre-commit hook to prevent broken versions from entering the project history, or as a check in a continuous integration (CI) pipeline to ensure that a manuscript is always ready for dissemination. This brings the rigour of professional software engineering to the academic writing process.

\section*{Discussion}

Rxiv-Maker sits alongside a growing set of tools for scientific publishing. It differs from collaborative web-based editors like Overleaf \cite{Overleaf2024}, which are strong for team-based LaTeX writing but do not keep a manuscript synchronised with external data and analysis code \cite{HenriquesLab2015_template}. It aligns more closely with platforms like Manubot \cite{himmelstein2019}, Jupyter Book \cite{JupyterBook2020}, and Quarto \cite{Quarto2024}, which also integrate code and narrative, often building on notebook-based formats \cite{Jupyter2016_notebook}. Rxiv-Maker is optimised for producing high-quality, submission-ready manuscripts from a local-first workflow. Beyond preprint distribution, it generates publication-quality PDFs and clean LaTeX source that many journals accept before their own formatting requirements, covering the path from initial draft through preprint distribution to journal submission. Its strengths (Table \ref{stable:tool-comparison}) are the self-updating capability, caching for rapid rebuilds, and deep integration with the Git-based workflows common in computational research. Other modern systems such as Typst \cite{Typst2024} and Bookdown \cite{Xie2016_bookdown} offer alternative approaches to scientific document creation.
Rxiv-Maker continues to evolve. Recent updates added Word export, letting authors share manuscripts with collaborators who use word processors. We plan to integrate Pandoc \cite{pandoc2020} more deeply for further output formats such as HTML and EPUB, while keeping typographic quality. We also intend to improve integration with computational environment managers to make executable manuscripts more portable and reproducible across systems \cite{biaflows2024,dl4miceverywhere2024}, for which containerisation is also a valid alternative \cite{Boettiger2015_docker_reproducibility,gomes2018}.
Rxiv-Maker applies the principles of literate programming \cite{Knuth1984_literate_programming} to academic papers, holding analysis code and text in one document so that a manuscript can be rebuilt from its sources and kept consistent with the analysis behind it.

\vspace{1em}

\begin{manuscriptinfo}
This work is licensed under CC BY 4.0.
\end{manuscriptinfo}

\begin{data}
The arXiv monthly submission data used in this article is available at \url{https://arxiv.org/stats/monthly_submissions}. Preprint submissions data across different hosting platforms is available at \url{https://github.com/esperr/pubmed-by-year}. The source code and data for the figures in this article are available at \url{https://github.com/HenriquesLab/rxiv-maker}.
\end{data}

\begin{code}
The Rxiv-Maker computational framework is available at \url{https://github.com/HenriquesLab/rxiv-maker}. The companion Visual Studio Code extension is at \url{https://github.com/HenriquesLab/vscode-rxiv-maker}. For users requiring containerised execution, the docker-rxiv-maker repository provides Docker-based deployment at \url{https://github.com/HenriquesLab/docker-rxiv-maker}. All repositories are released under an MIT License.
\end{code}

\begin{contributions}
Bruno M. Saraiva, Guillaume Jacquemet, and Ricardo Henriques conceived the project and designed the framework. António D. Brito beta-tested features and helped troubleshoot the user experience. Rita Carlota and Iván Hidalgo-Cenalmor tested features and helped edit the manuscript. All authors contributed to writing and reviewing the manuscript.
\end{contributions}

\begin{acknowledgements}
The authors thank Jeffrey Perkel for feedback that helped improve the manuscript.
\end{acknowledgements}

\begin{funding}
B.S. and R.H. acknowledge support from the European Research Council (ERC) under the European Union's Horizon 2020 research and innovation programme (SelfDriving4DSR, grant agreement No. 101001332 to R.H.) and funding from the European Union through Horizon Europe (AI4LIFE, grant agreement 101057970, and RT-SuperES, grant agreement 101099654, to R.H.). Funded by the European Union. However, the views and opinions expressed are those of the authors only and do not necessarily reflect those of the European Union. Neither the European Union nor the granting authority can be held responsible for them. R.H. also acknowledges a European Molecular Biology Organization (EMBO) Installation Grant (EMBO-2020-IG-4734), a Chan Zuckerberg Initiative Visual Proteomics Imaging award (vpi-0000000044, \url{https://doi.org/10.37921/743590vtudfp}), and a joint Wellcome, Chan Zuckerberg Initiative, and Kavli Foundation Essential Open Source Software for Science Cycle 6 award (Wellcome 313383/Z/24/Z; CZI EOSS6-0000000260). The Chan Zuckerberg Initiative awards are made through the Chan Zuckerberg Initiative DAF, an advised fund of Silicon Valley Community Foundation. A.D.B. acknowledges the FCT 2021.06849.BD fellowship. R.C. is supported by a doctoral contract funded by the "la Caixa" Foundation (CaixaResearch Health 2025, VirusAwareScopes, HR25-00453 to R.H.). I.H.C. is funded by the Finnish Doctoral Program Network in Artificial Intelligence (AI-DOC, decision number VN/3137/2024-OKM-6). This study was funded by the Research Council of Finland (338537, 371287, and 374180 to G.J.), the Sigrid Juselius Foundation (to G.J.), the Cancer Society of Finland (Syöpäjärjestöt, to G.J.), and the Solutions for Health strategic funding to Åbo Akademi University (to G.J.). G.J. is further supported by the InFLAMES Flagship of the Research Council of Finland (decision numbers 337530, 337531, and 357910) and by the Finnish Cancer Institute (K. Albin Johansson Professorship). This work was supported by FCT – Fundação para a Ciência e a Tecnologia, I.P., through the MOSTMICRO-ITQB R\&D Unit (DOI 10.54499/UID/04612/2025, UID/PRR/4612/2025) and the LS4FUTURE Associated Laboratory (DOI 10.54499/LA/P/0087/2020).
\end{funding}

\begin{interests}
The authors declare no competing interests.
\end{interests}

\begin{exauthor}
\begin{extendedauthorlist}
\extendedauthor{Bruno M. Saraiva}{\orcidicon{0000-0002-9151-5477}; \xicon{Bruno\_MSaraiva}; \linkedinicon{bruno-saraiva}}
\extendedauthor{Rita Carlota}{\orcidicon{0000-0003-3647-1114}}
\extendedauthor{António D. Brito}{\orcidicon{0009-0001-1769-2627}}
\extendedauthor{Iván Hidalgo-Cenalmor}{\orcidicon{0009-0000-8923-568X}}
\extendedauthor{Guillaume Jacquemet}{\orcidicon{0000-0002-9286-920X}; \twittericon{guijacquemet}; \blueskyicon{guijacquemet.bsky.social}}
\extendedauthor{Ricardo Henriques}{\orcidicon{0000-0002-2043-5234}; \xicon{HenriquesLab}; \blueskyicon{henriqueslab.bsky.social}; \linkedinicon{ricardo-henriques}}
\end{extendedauthorlist}
\end{exauthor}

\section*{Bibliography}
\bibliography{03_REFERENCES}

\section*{Methods}

The Rxiv-Maker framework is implemented as a modular, Python-based engine designed to ensure computational reproducibility at every stage of manuscript preparation. The processing pipeline begins by establishing an isolated and consistent computational environment, where all dependencies for Python, R, and LaTeX are managed. This ensures that a manuscript can be reliably recompiled on different machines and in the future.
The core of our approach is transforming the manuscript into an executable document. The engine identifies and runs embedded Python or R scripts to generate figures and statistical values directly from source data, as described in \ref{snote:programmatic_figures}. To keep authoring fast and interactive, this process is optimised with a caching system. It computes content-based checksums of the source scripts and their data dependencies, so that code is only re-executed when a genuine change has been made (\ref{snote:caching_validation}). This saves significant time during iterative writing without compromising the integrity of the results.
To convert the author's simple Markdown into a professional LaTeX document, we developed a multi-pass translator. This translator employs a content-protection strategy, first identifying and isolating delicate syntactic structures, including mathematical equations (\ref{snote:mathematical_formulas}), citation keys, and code blocks. Subsequent passes then normalise the document structure and map Rxiv-Maker's extended syntax for figures and cross-references to the appropriate LaTeX commands. This approach lets authors use the simplicity of Markdown without losing the syntax necessary for academic publishing.
A multi-level validation system provides continuous quality control. Before compilation, it checks for common errors, such as missing files or broken references. During compilation, it parses LaTeX and BibTeX logs to provide clear, actionable error messages. This automated quality control helps maintain the integrity of the document throughout the authoring process.
Finally, to ensure maximum reproducibility, the framework supports multiple deployment strategies. The primary recommended approach is a local installation, which requires Python and a LaTeX distribution. This method offers an excellent combination of high usability and reproducibility, with collaboration managed through Git. For absolute environmental consistency, which is particularly valuable for large collaborations or long-term archival, a fully containerised version is also available. This requires Docker but encapsulates the entire build environment, guaranteeing that the manuscript can be recompiled with bit-for-bit identical results on any system.

\onecolumn
\newpage



\renewcommand{\figurename}{Sup. Fig.}
\renewcommand{\tablename}{Sup. Table}
\setcounter{sfigure}{0}
\setcounter{stable}{0}

\newpage
\thispagestyle{empty}
\begin{center}

\vspace*{3cm}

\textbf{\Large Supplementary Information}

\vspace{3cm}

{\Huge\textbf{{\color{red}R}$\chi$iv-Maker: an automated template engine for streamlined scientific publications}}

\vspace{\fill}

\begin{minipage}{\textwidth}
\centering
\end{minipage}

\end{center}
\newpage

\begin{stable*}[ht]
\centering
\footnotesize
\begin{tabularx}{\textwidth}{|l|l|X|X|X|X|}
\hline
\textbf{Format} & \textbf{Input Extension} & \textbf{Processing Method} & \textbf{Output Formats} & \textbf{Quality} & \textbf{Use Case} \\
\hline
\textbf{Mermaid Diagrams} & \texttt{.mmd} & Mermaid CLI & SVG, PNG, PDF & Vector/Raster & Flowcharts, architectures \\
\hline
\textbf{Python and R Figures} & \texttt{.py}, \texttt{.R} & Script execution & PNG, PDF, SVG & Publication & Data visualisation \\
\hline
\textbf{Static Images} & \texttt{.png}, \texttt{.jpg}, \texttt{.svg} & Direct inclusion & Same format & Original & Photographs, logos \\
\hline
\textbf{LaTeX Graphics} & \texttt{.tex}, \texttt{.tikz} & LaTeX compilation & PDF & Vector & Mathematical diagrams \\
\hline
\textbf{Data Files} & \texttt{.csv}, \texttt{.json}, \texttt{.xlsx} & Python and R processing & Via scripts & Computed & Raw data integration \\
\hline
\end{tabularx}
\raggedright
\caption{\textbf{Supported Figure Generation Methods.} Overview of the framework's figure processing capabilities, demonstrating support for both static and dynamic content generation with an emphasis on reproducible computational graphics.}
\label{stable:figure_formats}
\end{stable*}

\begin{stable*}[ht]
\centering
\footnotesize
\begin{tabularx}{\textwidth}{|l|l|X|X|X|X|}
\hline
\textbf{Tool} & \textbf{Type} & \textbf{Markdown} & \textbf{Primary Use Case} & \textbf{Key Strengths} & \textbf{Open Source} \\
\hline
\textbf{Rxiv-Maker} & Pipeline & Excellent & Reproducible preprints & Local-first execution, automated caching, rich CLI & Yes \\
\hline
\textbf{Overleaf} \cite{Overleaf2024} & Web Editor & Limited & Collaborative LaTeX & Real-time collaboration, rich templates, cloud-based & Freemium \\
\hline
\textbf{Quarto} \cite{Quarto2024} & Publisher & Native & Multi-format publishing & Polyglot support, multiple outputs, scientific focus & Yes \\
\hline
\textbf{Manubot} \cite{himmelstein2019} & Collaborative & Native & Version-controlled writing & Automated citations, transparent collaboration, Git-based & Yes \\
\hline
\textbf{Pandoc} \cite{pandoc2020} & Converter & Excellent & Format conversion & Universal format support, extensible filters & Yes \\
\hline
\textbf{Typst} \cite{Typst2024} & Typesetter & Good & Modern typesetting & Fast compilation, modern syntax, growing ecosystem & Yes \\
\hline
\textbf{Bookdown} \cite{Xie2016_bookdown} & Publisher & R Markdown & Academic books & Cross-references, multiple formats & Yes \\
\hline
\textbf{Direct LaTeX} & Typesetter & None & Traditional publishing & Full control, established workflows, mature ecosystem & Yes \\
\hline
\end{tabularx}
\raggedright
\caption{\textbf{Comparison of Manuscript Preparation Tools.} This comparison positions each tool within the scientific publishing ecosystem. Rxiv-Maker specialises in reproducible preprint workflows with local-first execution and developer-centric features. Other tools address distinct needs, such as real-time collaborative editing (Overleaf), multi-format output (Quarto), or version-controlled writing (Manubot).}
\label{stable:tool-comparison}
\end{stable*}

\clearpage


\renewcommand{\thesubsection}{Supp. Note \arabic{subsection}}
\setcounter{subsection}{0}

\renewcommand{\thesubsection}{Supp. Note \arabic{subsection}}
\setcounter{subsection}{0}

\suppnotesection{A Getting Started Tutorial}\label{snote:getting_started}

To help new users begin with Rxiv-Maker, this tutorial provides a simple walkthrough for creating a minimal manuscript. The process starts with initialising a new project, which sets up the necessary directory structure and configuration files.

First, open a terminal and run the command \texttt{rxiv init my\_new\_paper}. This creates a new directory named \texttt{my\_new\_paper} containing the essential files: \texttt{00\_CONFIG.yml} for manuscript metadata, \texttt{01\_MAIN.md} for the main text, \texttt{02\_SUPPLEMENTARY\_INFO.md} for supplementary content, and \texttt{03\_REFERENCES.bib} for the bibliography. The system also creates subdirectories for \texttt{FIGURES} and \texttt{DATA}.

Next, you can edit the \texttt{01\_MAIN.md} file to add your text. You can use standard Markdown for formatting, such as \texttt{\detokenize{#}} for headings and \texttt{\detokenize{*}} for italics. To add a citation, simply use the \texttt{\detokenize{@key}} syntax, where \texttt{key} corresponds to an entry in your \texttt{.bib} file. For example, to cite the original literate programming paper, you would write \texttt{\detokenize{[@Knuth1984_literate_programming]}}.

To build the PDF, navigate into the project directory (\texttt{cd my\_new\_paper}) and run the command \texttt{rxiv pdf}. The framework will process your files, execute any embedded code, and generate a professionally typeset PDF in the \texttt{output} directory. If you encounter any issues, running \texttt{rxiv validate} provides a detailed check of your manuscript's integrity, flagging common problems like missing figures or broken citations.

This straightforward process allows you to get from a blank slate to a compiled PDF, providing a solid foundation that you can then build upon with more advanced features like programmatic figures and tables.

\suppnotesection{Rxiv-Maker Markdown Syntax and Advanced LaTeX Integration}\label{snote:markdown-syntax}

This reference shows the automated translation system that lets researchers write in familiar Markdown while producing professional LaTeX output. The table below covers the standard Markdown-to-LaTeX translations and is itself an example of Rxiv-Maker's \texttt{\detokenize{{{tex:...}}}} blocks, which allow direct LaTeX injection for advanced formatting.

For complex table structures that require precise control over formatting, multi-column headers, or mathematical notation, Rxiv-Maker's \texttt{\detokenize{{{tex:...}}}} syntax provides full access to LaTeX's typesetting capabilities. This table itself was created using \texttt{\detokenize{{{tex:...}}}} blocks, showing how raw LaTeX can be integrated into Markdown documents while the tex-block protection system prevents Markdown processing of LaTeX-specific syntax.

\small
\begin{longtable}{|p{0.28\textwidth}|p{0.3\textwidth}|p{0.34\textwidth}|}
\hline
\textbf{Markdown Input} & \textbf{LaTeX Output} & \textbf{Description} \\
\hline
\endfirsthead
\hline
\textbf{Markdown Input} & \textbf{LaTeX Output} & \textbf{Description} \\
\hline
\endhead

\multicolumn{3}{|c|}{\textbf{Basic Text Formatting}} \\
\hline
**bold text** & \textbackslash textbf\{bold text\} & Bold formatting \\
*italic text* & \textbackslash textit\{italic text\} & Italic formatting \\
\_\_underlined text\_\_ & \textbackslash underline\{underlined text\} & Underlined formatting for emphasis \\
**\_\_bold and underlined\_\_** & \textbackslash textbf\{\textbackslash underline\{bold and underlined\}\} & Nested formatting: bold containing underline \\
\_\_**underlined and bold**\_\_ & \textbackslash underline\{\textbackslash textbf\{underlined and bold\}\} & Nested formatting: underline containing bold \\
*\_\_italic and underlined\_\_* & \textbackslash textit\{\textbackslash underline\{italic and underlined\}\} & Multiple formatting combinations \\
\textasciitilde subscript\textasciitilde & \textbackslash textsubscript\{subscript\} & Subscript formatting, e.g., H\textsubscript{2}O, CO\textsubscript{2} \\
\textasciicircum superscript\textasciicircum & \textbackslash textsuperscript\{superscript\} & Superscript formatting, e.g., E=mc\textsuperscript{2}, x\textsuperscript{n} \\
\hline

\multicolumn{3}{|c|}{\textbf{Document Structure}} \\
\hline
\# Header 1 & \textbackslash section\{Header 1\} & Top-level section \\
\#\# Header 2 & \textbackslash subsection\{Header 2\} & Second-level section \\
\#\#\# Header 3 & \textbackslash subsubsection\{Header 3\} & Third-level section \\
\hline

\multicolumn{3}{|c|}{\textbf{Lists}} \\
\hline
- list item & \textbackslash begin\{itemize\}\textbackslash item...\textbackslash end\{itemize\} & Unordered list \\
1. list item & \textbackslash begin\{enumerate\}\textbackslash item...\textbackslash end\{enumerate\} & Ordered list \\
\hline

\multicolumn{3}{|c|}{\textbf{Links and URLs}} \\
\hline
{[}link text{]}(url) & \textbackslash href\{url\}\{link text\} & Hyperlink with custom text \\
https://example.com & \textbackslash url\{https://example.com\} & Bare URL \\
\hline

\multicolumn{3}{|c|}{\textbf{Citations}} \\
\hline
\texttt{@}mycitation & \textbackslash cite\{mycitation\} & Single citation \\
{[}@himmelstein2019;@Overleaf2024{]} & \textbackslash cite\{himmelstein2019,Overleaf2024\} & Multiple citations \\
\hline

\multicolumn{3}{|c|}{\textbf{Cross-References}} \\
\hline
figure:label & \textbackslash ref\{fig:label\} & Figure reference \\
suppfig:label & \textbackslash ref\{sfig:label\} & Supplementary figure \\
table:label & \textbackslash ref\{table:label\} & Table reference \\
suptab:label & \textbackslash ref\{stable:label\} & Supplementary table \\
equation:label & \textbackslash eqref\{eq:label\} & Equation reference \\
note:label & \textbackslash sidenote\{label\} & Supplement note reference \\
\hline

\multicolumn{3}{|c|}{\textbf{Tables and Figures}} \\
\hline
Markdown table & \textbackslash begin\{table\}...\textbackslash end\{table\} & Automatic table formatting \\
Image with caption & \textbackslash begin\{figure\}...\textbackslash end\{figure\} & Figure with caption \\
\hline

\multicolumn{3}{|c|}{\textbf{Document Control}} \\
\hline
\textless !-- comment --\textgreater & \% comment & LaTeX comments \\
<newpage> & \textbackslash newpage & Manual page break \\
<clearpage> & \textbackslash clearpage & Page break with float clearing \\
\hline
\caption{\textbf{Rxiv-Maker Markdown to LaTeX Translation Reference.} Mapping of Markdown syntax to corresponding LaTeX commands, showing the automated translation system that lets researchers write in familiar markup while producing professional typesetting.}
\end{longtable}

\suppnotesection{Programmatic Figure Generation}\label{snote:programmatic_figures}

Rxiv-Maker's figure generation keeps a transparent, reproducible connection between your data and your final visualisations. The system supports two main approaches for creating figures programmatically: script-based generation using Python or R, and diagram rendering from text-based descriptions using Mermaid.

For script-based figures, you place your \texttt{.py} or \texttt{.R} scripts in the \texttt{FIGURES} directory. These scripts often use plotting libraries such as Matplotlib \cite{Hunter2007_matplotlib} or Seaborn \cite{Waskom2021_seaborn}. During compilation, Rxiv-Maker executes these scripts, and any image files they save (e.g., PNG, PDF, SVG) are automatically detected and can be included in your manuscript. This ensures your visualisations are always synchronised with the underlying data and analysis, as a change in one will trigger the regeneration of the other. This is the method used to produce Fig. \ref{sfig:arxiv_growth} and Fig. \ref{sfig:preprint_trends}.

For diagrams, such as flowcharts or system architectures, you can use Mermaid \cite{Mermaid2023_documentation}. You create a \texttt{.mmd} file containing a text-based description of your diagram. The framework uses the Mermaid command-line tool to render this description into a vector or raster image. This allows your diagrams to be version-controlled just like code, making them easy to modify and track over time.

\suppnotesection{Mathematical Formula Support}\label{snote:mathematical_formulas}

Rxiv-Maker integrates mathematical notation by translating Markdown-style expressions into high-quality LaTeX mathematics. This allows you to write complex mathematical content using simple, familiar syntax.

For inline mathematics, you can use single dollar sign delimiters (\texttt{\detokenize{$...$}}), allowing formulas like $E = mc^2$ to be embedded directly within your text. For larger, display-style equations, you can use double dollar signs (\texttt{\detokenize{$$...$$}}) to centre the expression on its own line. For example:

$$
i\hbar\frac{\partial}{\partial t}\Psi(\mathbf{r},t) = \hat{H}\Psi(\mathbf{r},t)
$$

The framework's multi-pass translator protects these mathematical expressions so they are not altered during the conversion from Markdown to LaTeX. It supports mathematical and statistical notation, from simple symbols to multi-line equations.

\suppnotesection{Caching and Validation}\label{snote:caching_validation}

To speed up compilation, Rxiv-Maker uses a caching system that avoids redundant work. It generates a checksum for each figure's dependencies, including the source script and any input data files. A figure is only regenerated if this checksum changes, so recompilation stays fast during writing because only the modified components are rebuilt. This saves time while keeping the final document reproducible.

Alongside this is a validation framework for quality control. Running \texttt{rxiv validate} performs a multi-level check of your manuscript. Before compilation, it looks for missing figure files, broken cross-references, and malformed bibliography entries. During compilation, it parses LaTeX logs to provide clear, understandable error messages. After compilation, it even performs a lightweight scan of the PDF to flag potential rendering issues. This ensures your manuscript is technically sound at every stage.


\begin{sfigure*}[p!]
\centering
\makebox[\textwidth][c]{
  \includegraphics[width=0.700\textwidth,keepaspectratio,draft=false]{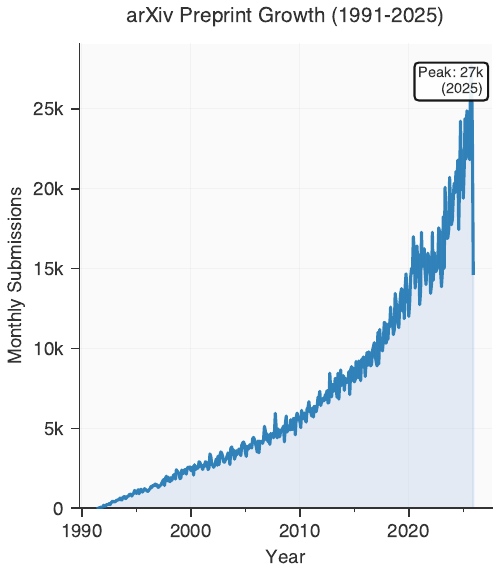}
}
\begingroup\captionsetup{width=0.95\textwidth,singlelinecheck=false,justification=justified}\setlength{\abovecaptionskip}{6pt}\setlength{\belowcaptionskip}{6pt}\ifdefined\justifying\justifying\fi\caption{\textbf{The growth of preprint submissions on the arXiv server (1991-2025).} This figure was generated from public arXiv statistics using a Python script executed by the Rxiv-Maker pipeline, demonstrating reproducible, data-driven visualisation.}\label{sfig:arxiv_growth}\endgroup
\end{sfigure*}

\begin{sfigure*}[p!]
\centering
\makebox[\textwidth][c]{
  \includegraphics[width=0.700\textwidth,keepaspectratio,draft=false]{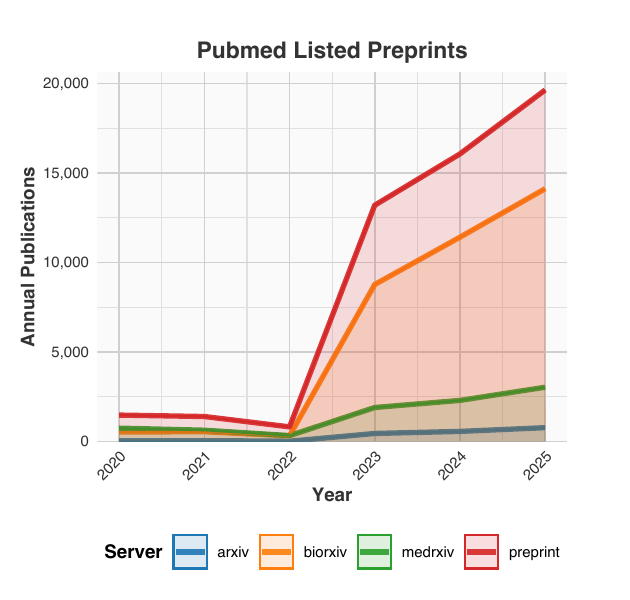}
}
\begingroup\captionsetup{width=0.95\textwidth,singlelinecheck=false,justification=justified}\setlength{\abovecaptionskip}{6pt}\setlength{\belowcaptionskip}{6pt}\ifdefined\justifying\justifying\fi\caption{\textbf{Preprint Submission Trends Across Multiple Servers (2018-2025).} This figure, showing preprints indexed by PubMed from major repositories, was generated from public data \cite{PubMedByYear2025} using a reproducible R script within the Rxiv-Maker pipeline.}\label{sfig:preprint_trends}\endgroup
\end{sfigure*}

\begin{sfigure*}[p!]
\centering
\makebox[\textwidth][c]{
  \includegraphics[width=1.000\textwidth,keepaspectratio,draft=false]{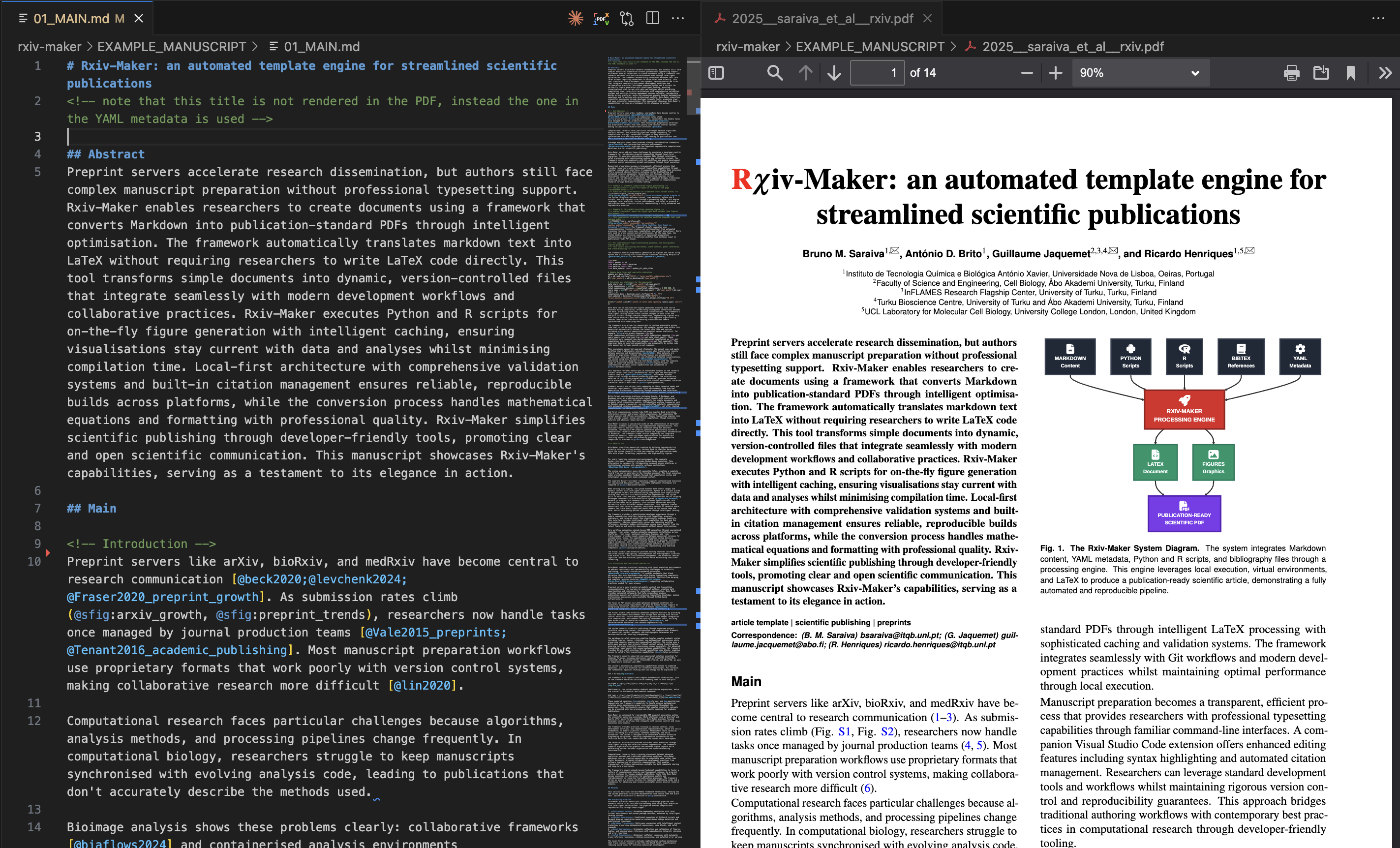}
}
\begingroup\captionsetup{width=0.95\textwidth,singlelinecheck=false,justification=justified}\setlength{\abovecaptionskip}{6pt}\setlength{\belowcaptionskip}{6pt}\ifdefined\justifying\justifying\fi\caption{\textbf{The Rxiv-Maker Visual Studio Code Extension.} The extension enhances the authoring experience by providing syntax highlighting for extended Markdown, autocompletion for citation keys (e.g., \texttt{\detokenize{@Knuth...}}), and real-time validation to catch errors as you type.}\label{sfig:vscode_extension}\endgroup
\end{sfigure*}

\end{document}